\documentclass[aps,pra,longbibliography,twocolumn,showpacs,superscriptaddress]{revtex4-1}
\usepackage{amsmath}
\usepackage{amssymb}
\usepackage{amsfonts}
\usepackage{txfonts}
\usepackage{bbold}
\usepackage{color}
\usepackage{bm}
\usepackage{graphicx}
\usepackage{epstopdf}
\usepackage{enumerate}

\usepackage[unicode,breaklinks]{hyperref}
\hypersetup{
    unicode=true,
    plainpages=false,
    pdftitle={Signatures of Coherent Vortex Structures in a Disordered 2D Quantum Fluid},
    pdfauthor={M. T. Reeves, T. P. Billam, B. P. Anderson, and A. S. Bradley},
    pdfsubject={Signatures of Coherent Vortex Structures in a Disordered 2D Quantum Fluid},
    colorlinks=true,
    linkcolor=blue,
    citecolor=blue,
    filecolor=black,
    urlcolor=blue
}
\usepackage{enumerate}
\newcommand {\eref}[1]{(\ref{#1})}

\newcommand{\bk}{{\mathbf k}}

\newcommand{\brr}{{\mathbf r}}

\newcommand{\EQ}[1]{\begin{eqnarray}#1\end{eqnarray}}

\newcommand{\rr}{\mathbf{r}}
\newcommand{\kk}{\mathbf{k}}
\newcommand{\yy}{\mathbf{y}}
\newcommand{\xx}{\mathbf{x}}
\newcommand{\va}{v}

\newcommand{\py}{\chi}   
\newcommand{\pxc}{{\sigma}}

\newcommand{\uu}{{\mathbf u}}
\newcommand{\vv}{{\mathbf v}}
\begin{document}
\title{Signatures of Coherent Vortex Structures in a Disordered 2D Quantum Fluid }
\author{Matthew T. Reeves}
\affiliation{Jack Dodd Center for Quantum Technology, Department of Physics, University of Otago, Dunedin, New Zealand}
\author{Thomas P. Billam}
\affiliation{Jack Dodd Center for Quantum Technology, Department of Physics, University of Otago, Dunedin, New Zealand}
\author{Brian P. Anderson}
\affiliation{College of Optical Sciences, University of Arizona, Tucson, Arizona 85721, USA}
\author{Ashton S. Bradley} 
\affiliation{Jack Dodd Center for Quantum Technology, Department of Physics, University of Otago, Dunedin, New Zealand}
\date{\today}
\begin{abstract}
The emergence of coherent rotating structures is a phenomenon characteristic of
both classical and quantum 2D turbulence.  In this work we show theoretically
that the coherent vortex structures that emerge in decaying 2D quantum
turbulence can approach quasi-classical rigid-body rotation, obeying the
Feynman rule of constant average areal vortex density while remaining spatially
disordered. By developing a rigorous link between the velocity probability
distribution and the quantum kinetic energy spectrum over wavenumber $k$, we
show that the coherent vortex structures are associated with a $k^3$ power law
in the infrared region of the spectrum, and a well-defined spectral peak that
is a physical manifestation of the largest structures. We discuss the possibility of realizing coherent structures in Bose--Einstein condensate experiments and present Gross-Pitaevskii simulations showing that this phenomenon, and its associated spectral signatures, can emerge dynamically from feasible initial vortex configurations.

\end{abstract}
\maketitle
\section{Introduction}
The emergence of coherent rotating structures from disordered flows is a
central feature of 2D classical turbulence
\cite{Kra1980.RPP5.547,Tabeling2002,Boffetta12a}. In two-dimensional quantum
turbulence
(2DQT)~\cite{Numasato10a,Numasato10b,Bradley2012a,Reeves12a,neely_etal_prl_2013,Wilson2013,White2013a},
the analogous phenomenon involves large-scale clustering of quantum vortices of
the same sign of circulation; such clustering can occur in negative temperature
equilibrium states~\cite{Onsager1949,Eyi2006.RMP78.87,Billam2013a}, and as a
result of a turbulent inverse-energy cascade induced by small-scale
forcing~\cite{reeves_etal_prl_2013}. The
characterization of such clustered vortex states, which tend to be highly
disordered arrangements of same-sign vortices, poses a theoretical challenge of
recent interest \cite{White2012a,Reeves12a}.  In many respects, these states
strongly contrast with a rotating superfluid in its ground state, which will
form a regular Abrikosov lattice comprised of co-rotating vortices, exhibiting
a sixfold rotational symmetry, and obeying Feynman's rule of constant areal
vortex density \cite{Feynman1955,Fetter09a}.  Although the velocity field of
the superfluid is formally curl-free, the velocity field of a large lattice
approaches that of rigid-body rotation under appropriate coarse-graining over
many vortex cores~\cite{Fetter2001}, as required by Bohr's correspondence
principle. Furthermore, the self-similar expansion of the atomic density of a
3D turbulent cloud~\cite{Henn09a} can be modelled by introducing a rotational
velocity field~\cite{Caracanhas2012a}, suggesting that the development of a
\emph{rotational} velocity field in an \emph{irrotational} superfluid may be a
fundamental property of quantum turbulence. In the context of the negative
temperature states arising as the end states of decaying 2DQT \cite{Billam2013a}, these
considerations motivate the question: \emph{what kind of rotational velocity
field emerges in the interior of a large, coherent vortex cluster?} 

In this work we address the problem of characterizing emergent coherent vortex structures in a 2D Bose-Einstein condensate (BEC), a compressible superfluid, with an emphasis on experimentally accessible measures.
While measurements of vortex locations and circulations~\cite{White2012a,reeves_etal_prl_2013}, or of two-point velocity correlations would allow inference of the \textit{classical} (point-vortex-like) energy spectrum~\cite{Bradley2012a, Nore97a}, neither the vortex circulations, nor the velocity correlations are easily accessed in current 2DQT experiments \cite{Wilson2013,neely_etal_prl_2013}. These difficulties motivate an exploration of other measures that may be more easily accessible.  
Here, we consider information contained in the quantum kinetic energy spectrum of the quantum fluid; provided interactions can be suppressed using, for example, an appropriate Feshbach resonance, this information may be readily available in experiments via ballistic expansion \cite{Thompson2014a}. 
 We develop a link between the quantum kinetic energy spectrum over wavenumber $k$ and the superfluid velocity probability distribution, and show analytically that a spectrum $E(k)\propto k^3$  in the infrared arises from the coherent quantum vortex structures emerging in negative-temperature vortex configurations. We show analytically that  such a spectrum can correspond to rigid-body rotation, extending to a scale determined by the size of the coherent vortex structures, and, additionally, can arise due to quadrupole velocity fields resulting from the interaction between the coherent structures and the superfluid boundary. 
 
We numerically sample vortex configurations over a range of energies, exploring
a number of measures with which to characterize the vortex distributions. We
verify the emergence of a $k^3$ spectrum and demonstrate that the coherent
structures produce a well-defined peak in the quantum kinetic energy spectrum.
This peak can be quantitatively understood by considering the largest vortex
structures. We further find that the largest structures exhibit a constant
areal vortex density, and conclude that the clustered states that emerge in
negative-temperature configurations obey Feynman's rule while being spatially
disordered. We consider the outlook for observing quasi-classical coherent
vortex structures in atomic BEC experiments. Dynamical simulations of a trapped
BEC (within the damped Gross-Pitaevskii description) show that, for
well-chosen non-equilibrium vortex configurations that may be accessible via
laser-stirring protocols~\cite{Neely10a}, the resulting vortex dynamics can
form long-lived, high-energy coherent vortex structures despite some loss of
energy to sound.

Our main result is that quasi-classical coherent structures exhibiting rigid-body rotation can emerge in negative-temperature vortex states, establishing a new link between classical and quantum turbulence that may be explored experimentally. 

This paper is structured as follows. In Sec. \ref{sec:Background} we introduce relevant background, and the classical spectral decomposition frequently used in the literature. In Sec. \ref{sec:KineticnEnergySpectrum} we develop a decomposition of the quantum kinetic energy spectrum of the quantum fluid, and show its connection to the velocity probability distribution for a compressible superfluid in the hydrodynamic regime. 
In Sec. \ref{sec:NumericalSampling} we present our numerical methods for
sampling and analyzing clustered states, and analyze kinetic energy spectra of
vortex distributions over a range of energies for vortex configurations in a
doubly-periodic box. In Sec. \ref{sec:ClassicalFlows} we analytically and
numerically investigate the emergence of quasi-classical flows for
negative-temperature states. In Sec. \ref{sec:Trapped} we show that high-energy coherent vortex structures, and the associated spectral signatures, can emerge dynamically in a trapped BEC, and compare the properties of the dynamically-generated structures to the properties predicted by microcanonical sampling.
Sec. \ref{sec:Conclusion} presents concluding remarks.  
\section{Background}
\label{sec:Background}
We consider a BEC that is tightly confined in the $z$-direction. The Gross-Pitaevskii equation describing this homogeneous  2D Bose gas is written in terms of an effective 2D interaction parameter $g_2$:
\EQ{\label{GPEdef}
i\hbar\frac{\partial \psi(\brr,t)}{\partial t}&=&\left(-\frac{\hbar^2\nabla_\perp^2}{2m}+g_{2}|\psi(\brr,t)|^2\right)\psi(\brr,t),
}
where $g_2= g/l$, $l$ is the characteristic thickness of the 3D system \cite{Numasato10b}, and $g=4\pi\hbar^2 a_s/m$ for $s$-wave scattering length $a_s$ and atomic mass $m$.
For example, in a system with harmonic trapping in the $z$-direction characterized by trapping frequency $\omega_z$, the length scale is $l=\sqrt{2\pi}l_z$ where $l_z=\sqrt{\hbar/m\omega_z}$ is the $z$-axis harmonic oscillator length, and the confinement is assumed sufficient to put the wavefunction into the $z$-direction single-particle harmonic oscillator ground state. We note, however, that such tight confinement in one direction is not necessarily required for studies of 2DQT~\cite{Rooney11a,neely_etal_prl_2013}.
\subsection{Properties of a 2D  Quantum Vortex}
For solutions with chemical potential $\mu$ containing a single vortex at the origin (with circulation necessarily normal to the plane of the 2D quantum fluid) we can write~\cite{Fetter2001}
\EQ{\label{oneVdef}
\psi_1(\brr,t)=\sqrt{n_0}e^{-i\mu t/\hbar}\chi\!\left(r/\xi\right)e^{\pm i\theta}
}
where $\xi=\hbar/m c$ is the healing length for speed of sound $c=\sqrt{\mu/m}$, and $n_0=\mu/g_2$ is the 2D particle density for $r\gg \xi$ and is taken to be a constant. The vortex radial amplitude function $\py(\pxc)$, where $\pxc = r/\xi$ is a scaled radial coordinate, is a solution of 
\EQ{\label{yeq}
\left(-\pxc^{-1}\partial_\pxc\, \pxc\partial_\pxc+\pxc^{-2}\right)\py=2(\py-\py^3).
}
The boundary conditions are $\py(0)=0$, and the derivative $\py^{\prime} \equiv d \py/d\pxc$ evaluated at $\pxc=0$ must be chosen such that it is consistent with $\py(\infty)=1$ and $\py^\prime(\infty)=0$. The value 
\EQ{\label{coreDeriv}
\Lambda\equiv \py^\prime(0)=\lim_{r\to 0}\frac{\xi}{\sqrt{n_0}}\left|\frac{d\psi_1}{dr}\right|
}
is a universal feature of the vortex core, and numerically is found to be  $\Lambda =0.8249\dots$ \cite{Bradley2012a}. The quantum vortex state (\ref{oneVdef}) has the velocity field of a point-vortex 
\EQ{\label{vortV}
\mathbf{v}(\brr)=\frac{\hbar}{mr}(\mp\sin\theta,\pm\cos\theta)=(v_x,v_y),
}
which has vorticity, $\omega(\brr) = \partial_x v_y-\partial_yv_x$, given by 
\begin{equation}
\label{vorticity}
\omega(\rr) = \pm\frac{h}{m}\delta(\rr),
\end{equation}
where $\delta(\rr)$ is the Dirac $\delta$-function.
\subsection{Hydrodynamic decomposition}
The 2D Gross-Pitaevskii energy in the homogeneous system is given by
\begin{equation}
\mathcal{E} = \int d^2\rr \left\{ \frac{\hbar^2}{2m}|\nabla\psi(\rr,t)|^2 +  \frac{g_2}{2}|\psi(\rr,t)|^4 \right\}.
\end{equation}
Using the Madelung representation $\psi(\rr,t)=\sqrt{\rho(\rr,t)}e^{i\theta(\rr,t)}$ which gives the superfluid velocity as $\vv(\rr,t) = \hbar\nabla\theta(\rr,t)/m$, the energy can be decomposed as $\mathcal{E}=\mathcal{E}_{H}+\mathcal{E}_Q+\mathcal{E}_{I}$, where
\EQ{\label{EtermsK}
\mathcal{E}_H&=&\frac{m}{2}\int d^2\brr\; \rho(\brr,t)|\mathbf{v}(\brr,t)|^2,\\
\label{EtermsQ}
\mathcal{E}_Q&=&\frac{\hbar^2}{2m}\int d^2\brr\; |\nabla\!\sqrt{\rho(\brr,t)}|^2,\\
\label{EtermsI}
\mathcal{E}_I&=&\frac{g_2}{2}\int d^2\brr\; \rho(\brr,t)^2.
}
Respectively, these define the hydrodynamic kinetic energy, quantum pressure energy, and interaction energy. The hydrodynamic and quantum pressure terms originate from the kinetic energy term, and  $\mathcal{E}_K = \mathcal{E}_H + \mathcal{E}_Q$ is the total kinetic energy. 
\subsection{Classical Kinetic Energy Spectrum}

It is worthwhile to  briefly review the spectral decomposition often used in the literature, for example Ref. \cite{Nore97a}. We call this the \emph{classical} kinetic energy spectrum, as in  a classical fluid it is exactly the kinetic energy power spectrum. This spectrum is obtained by applying the general correspondence between a two-point correlation function and its associated power spectrum to the velocity field. However, as we will show in Sec.~\ref{sec:KineticnEnergySpectrum}, in a quantum fluid the existence of a quantum phase $\theta(\rr)$ breaks this correspondence. While in a quantum fluid this classical spectrum is no longer the kinetic energy spectrum, it provides a useful link to classical turbulence theory, allowing the identification of, for example, the Kolmogorov $k^{-5/3}$ law associated with an inertial range. 

As we will only focus on particular instants in time, we now drop the explicit time dependence from our notation.  By Parseval's theorem, \eref{EtermsK}  may be equivalently written in wavenumber $(k)$-space as
\EQ{\label{ikek}
\mathcal{E}_H=\frac{m}{2}\int d^2\bk\;|\tilde{\uu}(\bk)|^2,
}
where
\EQ{\label{Fj}
\tilde{\uu}(\bk) = \mathcal{F}[\uu(\rr)] \equiv  \frac{1}{2\pi}\int d^2\brr\; e^{-i\bk\cdot \brr}\mathbf{u}(\brr),
}
and $\uu(\rr) = \sqrt{\rho(\rr)}\vv(\rr)$ is the density-weighted velocity field.
The one-dimensional \emph{spectral density} in $k$-space is given in polar coordinates by integrating over the azimuthal angle to give

\begin{eqnarray}
\epsilon_H(k)&=&\frac{mk}{2} \int_0^{2\pi}d\theta_k \;|\tilde{\uu}(\bk)|^2 \label{ikeks}\\
& = & \frac{mk}{2} \int_0^{2\pi} d\theta_k \; \left|\frac{1}{2\pi} \int d^2\rr \; e^{-i\mathbf{k} \cdot \rr}\uu(\rr)\right|^2 \label{ikeks2}
\end{eqnarray}
which, when integrated over all $k$, gives the total hydrodynamic kinetic energy $\mathcal{E}_H=\int_0^\infty dk\; \epsilon_H(k)$. Similarly, for the quantum pressure we have
\EQ{\label{EQk}
\epsilon_Q(k)\equiv \frac{\hbar^2}{2m}k\int_0^{2\pi}d\theta_k\left|\frac{1}{2\pi}\int d^2\rr\; e^{-i\kk\cdot\rr}\nabla \sqrt{\rho(\rr)}\right|^2,
}
and  the total kinetic energy is given by   $\mathcal{E}_K = \mathcal{E}_H + \mathcal{E}_Q =  \int dk \: [ \epsilon_H(k) + \epsilon_Q(k) ] $.

 The hydrodynamic spectrum $\epsilon_H(k)$ can be further decomposed into incompressible and compressible parts via a Helmholtz decomposition, writing $\uu = \uu^i + \uu^c$, where $\nabla \cdot \uu^i = 0$ and $\nabla \times \uu^c = 0$. The incompressible part is associated with quantum vortices, whereas the compressible part is associated with acoustic excitations \cite{Numasato10b,Bradley2012a}. Although we do not make direct use of this decomposition here,  we are generally interested in the incompressible limit, where $\uu \approx \uu^i$.  As a single vortex is purely incompressible~\cite{Bradley2012a}, the incompressible limit corresponds to a fluid for which the background density is smoothly varying, and the quantum vortices are sufficiently well-separated that their cores do not overlap.

It has become standard in the literature to interpret the incompressible and compressible parts of $\epsilon_H(k)$ loosely as  kinetic energy densities in $k$-space~\cite{reeves_etal_prl_2013,Kobayashi05a,Numasato10b}, as is the case for classical fluids, and thus these kinetic energy densities are generally referred to as ``incompressible" and ``compressible" spectra. However, this approach does not provide a true kinetic energy spectrum that can be directly connected to the momentum distribution of a quantum fluid. Furthermore, the classical ``spectra" are not locally additive in $k$-space, complicating the identification of energy fluxes.

\section{Kinetic Energy Spectrum and Velocity Probability Distribution}\label{sec:KineticnEnergySpectrum}
In this section we pursue an alternative route to decomposing the kinetic energy of the quantum fluid, and develop a link between the true quantum kinetic energy spectrum and the velocity probability distribution.

\subsection{Quantum Kinetic Energy Spectrum}

The kinetic energy of the quantum fluid is given by
\EQ{\label{ketot}
\mathcal{E}_{K}=\frac{\hbar^2}{2m}\int d^2\rr \;|\nabla\psi(\rr)|^2.
}
This may be equivalently written in momentum space as
\EQ{\label{kspace}
\mathcal{E}_{K}=\frac{\hbar^2}{2m}\int d^2\kk\;|\kk\phi(\kk)|^2,
}
where 
\EQ{\label{phi}
\phi(\kk)  = \frac{1}{2\pi}\int d^2\rr\; e^{-i\mathbf{k}\cdot \rr} \psi(\rr).
}
Writing $\kk=k(\cos\theta_k,\sin\theta_k)$, and performing the angular integration, we obtain the \emph{quantum} kinetic energy spectrum
\begin{equation}\label{Ekquant}
E(k) = \frac{\hbar^2k^3}{2m} \int_0^{2\pi} d\theta_k \; |\phi(\mathbf{k})|^2,
\end{equation}
and the total kinetic energy via $\mathcal{E}_K = \int_0^{\infty} dk \; E(k)$.

We now provide a decomposition of the quantum kinetic energy spectrum that allows the identification of hydrodynamic and quantum pressure components, and their relationship to the momentum distribution. Returning to Eq.~\eref{kspace}, the integrand can  be decomposed using \eref{phi} and the Madelung transformation. 

We thus write the total kinetic energy as
\EQ{\label{Ekkdef}
\mathcal{E}_{K}&=&\int_0^\infty dk\;E(k)\\
&=&\int_0^\infty dk\; \left[E_H(k)+E_Q(k)+E_{QH}(k)\right],\label{Ekqdef}
}
where 
\EQ{\label{epsH}
E_H(k)&=&\frac{mk}{2}\int_0^{2\pi}d\theta_k\left|\frac{1}{2\pi}\int d^2\rr\; e^{-i\kk\cdot\rr+i\theta(\rr)}\mathbf{u}(\rr)\right|^2,\\
\label{epsQ}
E_Q(k)&=&\frac{\hbar^2k}{2m}\int_0^{2\pi}d\theta_k\left|\frac{1}{2\pi}\int d^2\rr\; e^{-i\kk\cdot\rr+i\theta(\rr)}\nabla \sqrt{\rho(\rr)}\right|^2,\;\;\;\\
\label{epsf}
E_{QH}(k)&=&\frac{\hbar k}{2}\int_0^{2\pi}d\theta_k\Phi(\kk),
}
and
\EQ{\label{ferr}
\Phi(\kk)&=&\frac{-i}{(2\pi)^2}\int d^2\rr e^{-i\kk\cdot \rr+i\theta(\rr)}\nabla\sqrt{\rho(\rr)}\nonumber\\
&&\cdot\int d^2\rr^\prime e^{i\kk\cdot \rr^\prime-i\theta(\rr^\prime)}\mathbf{u}(\rr^\prime) \nonumber\\
&&+\mathrm{c.c.},
}
with $\mathrm{c.c.}$ denoting the complex conjugate. Equations \eref{epsH}, \eref{epsQ}, and \eref{epsf} 
give the kinetic energy spectra for the hydrodynamic, quantum pressure, and quantum-hydrodynamic components respectively. 

 The decomposition derived above provides genuine spectral energy densities, as they are locally additive in $k$-space: $E_H(k) + E_Q(k) + E_{QH}(k) = E(k)$. In contrast, although \emph{integrating} the classical spectra yields $\mathcal{E}_H+\mathcal{E}_Q=\mathcal{E}_{K}$, the classical spectra are not locally additive in $k$-space: $\epsilon_H(k) + \epsilon_Q(k) \neq E(k)$. The hydrodynamic and quantum pressure terms here differ to those of the classical spectral decomposition by the formal replacement $\mathbf{u}(\rr)\rightarrow e^{i\theta(\rr)} \mathbf{u}(\rr)$ from \eref{ikeks2} to \eref{epsH} and $\nabla\sqrt{\rho(\rr)}\rightarrow e^{i\theta(\rr)}\nabla\sqrt{\rho(\rr)}$ from \eref{EQk} to  \eref{epsQ}.   Notice also that there is no term corresponding to the quantum-hydrodynamic term \eref{epsf} in the classical kinetic energy decomposition. Furthermore, as 
\begin{multline}
\mathcal{E}_K = \mathcal{E}_H + \mathcal{E}_Q =\int_0^\infty dk \; [ \epsilon_H(k) + \epsilon_Q(k)] \\ 
 = \int_0^\infty dk\; [ E_H(k) + E_Q(k) + E_{QH}(k)],
\end{multline}
we may conclude that
\begin{equation}
\int_0^\infty dk \; E_{QH}(k) = 0,
\end{equation}
illustrating that the quantum-hydrodynamic term does not contribute to the total energy, and serves only to redistribute the energy in $k$-space. 
Another important property of the quantum kinetic energy
spectrum [Eq.~(19)] is its potential experimental accessibility: The momentum
distribution of a non-interacting condensate, $|\phi(\mathbf{k})|^2$, can be
obtained through ballistic (time-of-flight) expansion \cite{Thompson2014a, Dalfovo1999}.
Thus, after suppressing interatomic interactions via an appropriate Feshbach
resonance, one can obtain $E(k)$ to high accuracy.

The spectral decomposition we have introduced here thus highlights an important distinction between classical and quantum fluids. In a quantum fluid Eq. \eref{epsH} is the true  hydrodynamic kinetic energy spectrum, whereas Eq. \eref{ikeks} is  the power spectrum of the velocity autocorrelation function.  In a classical fluid there is no quantum phase, and hence there is no distinction between these two measures.  We remark that the quantum spectra may resemble the classical spectra in regimes where the phase $\theta(\rr)$ is approximately constant over large regions of the system, for example in the vortex-dipole gas regime~\cite{CheslerScience2013} (see also Sec.~\ref{sec:NumericalSampling}). 
\subsection{Velocity Probability Distribution}

The probability, $P(v)$, that an atom has velocity $v=|\vv|$ is
\EQ{\label{Pv}
P(v)=\frac{1}{N_{\rm{tot}}}\int d^2\rr\;\delta(v-|\vv(\rr)|)|\psi(\rr)|^2,
}
for $N_{\rm{tot}}=\int d^2\rr\; |\psi(\rr)|^2$ atoms, where the normalization is $\int_0^\infty dv\;P(v)=1$. Note that binning the velocity neglects the density weighting, and so in the context of 2DQT is equivalent to the above for a homogenous superfluid with coreless vortices~\cite{White10a,Paoletti08a}. The physical distinction is important in regions where the atomic density is rapidly varying, such as near a quantum vortex core. 

Let us briefly consider  the velocity probability distribution of a single vortex in a homogeneous, compressible superfluid. We start by calculating $P(\va)$ for atoms in a superfluid vortex, given by \eref{oneVdef}, in an otherwise homogeneous 2D system. Using \eref{oneVdef}, \eref{vortV}, and \eref{Pv}, and exploiting cylindrical symmetry, yields
\EQ{\label{vdist}
P(\va)=\frac{2\pi n_0\xi^2}{cN_{\rm{tot}}}\left(\frac{c}{v}\right)^3\left|\chi\left(\frac{c}{v}\right)\right|^2H(v/v_R),
}
where we have introduced the system size $R$, and the Heaviside function $H(v/v_R)$ restricts the range of velocities to $\va\geq v_R\equiv c\xi/R=\hbar/mR$, avoiding infrared divergence. 
 Two regimes can be identified within \eref{vdist}, namely a \emph{point-vortex regime} for $v_R \leq v \ll c$ (where $r \gg \xi$ )
\EQ{\label{Ppv}
P(\va)\Big{|}_{v\ll c}\simeq\frac{\xi^2}{R^2}\frac{2}{c}\left(\frac{c}{\va}\right)^3,
}
and a \emph{vortex core regime} for $v \gg c$ ($r\ll \xi $)
\EQ{\label{Pcore}
P(\va)\Big{|}_{v\gg c}\simeq\frac{\xi^2}{R^2}\frac{2\Lambda^2}{c}\left(\frac{c}{\va}\right)^5.
}
Notice the parameter $\Lambda$ [Eq. \eref{coreDeriv}] appears in the vortex core regime,  whereas this is absent in \eref{Ppv}. The $\va^{-5}$ power law seen here stems from the structure of a vortex core in an atomic Bose-gas superfluid. In a macroscopically occupied BEC with small healing length, the vortex core region only corresponds to a tiny fraction of the atoms. However, it might be possible to observe the $\va^{-5}$ power law when the system contains many vortices and few particles, namely, in the vortex-liquid phase~\cite{Fetter09a}.

 The $v^{-3}$ power-law tail of Eq.~\eref{Ppv} is a universal result for quantum vortices~\cite{Paoletti2011a}, and has been identified in the 2D and 3D GPE \cite{White10a}, the point-vortex model in 2D \cite{Min1996a,Weiss1998a}, and in superfluid helium \cite{Paoletti08a,Adachi2011a}, which is well-described by a vortex-filament model at larger scales. As the $v^{-3}$ power-law is a \emph{single vortex effect}~\cite{Paoletti08a,White10a,Paoletti2011a}), it will be present in, but is not indicative of, turbulent vortex dynamics. For scales smaller than the minimum inter-vortex distance but appreciably larger than the healing length, the single-vortex velocity field can be expected to dominate the distribution. However, effects due to cooperative behaviour of many quantum vortices are central to quantum turbulence~\cite{Bradley2012a,Billam2013a}, and may lead to different behavior for scales greater than the minimum inter-vortex distance \cite{Baggaley2011a,Adachi2011a}.

\subsection{Quantum Kinetic Energy Spectrum in the Hydrodynamic Regime}\label{sec:hydro}
In this subsection we evaluate the definition of the spectrum \eref{epsH} within a hydrodynamic approximation that neglects high-order density and phase gradients. This approximation gives a rigorous link between the velocity probability distribution and the quantum kinetic energy spectrum, applicable for a system of quantum vortices in a smoothly varying background density, while neglecting the density variations occurring within distance $\xi$ of a vortex core. 

We confine our attention to the hydrodynamic kinetic energy spectrum $E_H(k)$. Performing the integral over $\theta_k$, the spectrum \eref{epsH} becomes
\EQ{\label{spechyd}
E_H(k)&=&\frac{mk}{2}\frac{1}{2\pi}\int d^2\rr\int d^2\rr^\prime e^{i\left[\theta(\rr)-\theta(\rr^\prime)\right]}\nonumber\\
&&\times\sqrt{\rho(\rr)\rho(\rr^\prime)}\vv(\rr)\cdot\vv(\rr^\prime)J_0(k|\rr-\rr^\prime|).
}
Transforming to coordinates $\xx=(\rr+\rr^\prime)/2$, and $\yy=\rr-\rr^\prime$, we Taylor expand in powers of $\yy$ to give
\EQ{\label{expand}
\theta(\xx+\yy/2)-\theta(\xx-\yy/2)&\approx& \yy\cdot\nabla\theta(\xx)=\frac{m}{\hbar}\yy\cdot \vv(\xx),\\
\sqrt{\rho(\xx+\yy/2)}\sqrt{\rho(\xx-\yy/2)}&\approx&\rho(\xx)-\frac{1}{4}\left(\yy\cdot\nabla\sqrt{\rho(\xx)}\right)^2,\\
\vv(\xx+\yy/2)\cdot\vv(\xx-\yy/2)&\approx&|\vv(\xx)|^2-\frac{1}{4}\left|(\yy\cdot\nabla)\vv(\xx)\right|^2,
}
yielding the expression
\EQ{\label{EHk2}
E_H(k)&\approx&\frac{mk}{2}\frac{1}{2\pi}\int d^2\xx\int d^2\yy\; e^{im \vv(\xx)\cdot \yy/\hbar}J_0(ky)\nonumber\\
&&\times\Bigg[\rho(\xx)|\vv(\xx)|^2\label{order0}\\
&&-\frac{\rho(\xx)}{4}\big|(\yy\cdot\nabla)\vv(\xx)\big|^2\label{order1a}-\frac{|\vv(\xx)|^2}{4}\left(\yy\cdot\nabla\sqrt{\rho(\xx)}\right)^2\label{order1b}\\
&&+\frac{1}{16}\left(\yy\cdot\nabla\sqrt{\rho(\xx)}\right)^2\big|
(\yy\cdot\nabla)\vv(\xx)\big|^2\label{order2}\Bigg].
}
We may write this as $E_H(k) \approx E_H^{(0)}(k) + E_H^{(2)}(k) + E_H^{(4)}(k)$,  where the superscripts denote the orders of $\nabla$ involved in each term. In this paper we take a hydrodynamic approach and treat the lowest order term, validating the results against a full numerical treatment. 
Considering the lowest order term, Eq. \eref{EHk2}, and performing the angular integral in $\yy$ gives
\EQ{\label{EH0}
{E}_{H}^{(0)}(k)&=&\frac{mk}{2}\frac{1}{2\pi}\int d^2\xx\;\rho(\xx)|\vv(\xx)|^2\nonumber\\
&&\times\int_0^\infty y dy\;J_0\left(\frac{ym|\vv(\xx)|}{\hbar}\right)J_0(ky).
}
This can be evaluated using the Bessel closure relation $\int_0^\infty x dx\;J_\alpha(ux)J_\alpha(vx)=\delta(u-v)/u$, to give the  useful result 
\EQ{\label{EH0x}
E_H^{(0)}(k)&=&\frac{m}{2}\int d^2\rr\;\rho(\rr)|\vv(\rr)|^2\delta\left(k-\frac{m|\vv(\rr)|}{\hbar}\right).
}
 At this level of approximation, the kinetic energy is a simple local transform of the hydrodynamic kinetic energy. 

Given the above, let us note that integrating over $k$ gives
\EQ{\label{intEH}
\int_0^\infty dk\;E_H^{(0)}(k)=\frac{m}{2}\int d^2\rr\;\rho(\rr)|\vv(\rr)|^2\equiv \mathcal{E}_H,
}
and our approximations have not affected the total energy. Furthermore, using \eref{Pv} with \eref{EH0x}, we have
\EQ{\label{EHtildev}
E_H^{(0)}(k)=N_{\rm{tot}}\frac{\hbar}{2}\left(\frac{\hbar k}{m}\right)^2P\left(\frac{\hbar k}{m}\right),
}
where the total hydrodynamic kinetic energy is given by
\EQ{\label{KEItot}
\int_0^\infty E_H^{(0)}(k)dk&=&N_{\rm{tot}}\int_0^\infty dv\; P(v)\frac{mv^2}{2}.
}
The expression \eref{EHtildev} is our main analytical result, providing a rigorous link between the velocity distribution and the kinetic energy spectrum in the hydrodynamic regime. A superfluid state comprised of quantized vortices in a homogeneous background will be very well described by this expression for the kinetic energy in the regime $k\ll \xi^{-1}$, provided $\mathcal{E}_H$ is the dominant energy contribution, and the system does not contain a significant amount of acoustic energy. 

Eq. \eref{EHtildev} further emphasizes the difference between the classical and quantum kinetic energy spectra in a quantum fluid. The classical spectrum, Eq.~\eref{ikeks}, contains information about the \emph{spatial structure} of velocity correlations, but it is not a true kinetic energy spectrum for quantum fluids. Nevertheless, it is the classical spectrum of a quantum fluid which is actually analogous to the kinetic energy spectrum of classical fluids, the incompressible part of which can exhibit, for example, the Kolmogorov scaling \cite{Kobayashi05a,reeves_etal_prl_2013,Kobyakov2014a}, and spectral condensation \cite{Billam2013a}.  In contrast, whilst the quantum kinetic energy spectrum, Eq.~\eref{epsH}, draws information from the velocity \emph{probability  distribution} [as shown in Eq. \eref{EHtildev}], it has no obvious classical counterpart.

\subsection{Spectral Signatures of Coherent Structure Formation}
\label{Signatures}
For systems containing many vortices and little compressible energy, the quantum kinetic energy spectrum should be well described by the hydrodynamic approximation Eq. \eref{EHtildev}. We now consider the spectral features that may be observed for systems containing coherent vortex structures.  

A fundamental property of superfluidity is the constraint that the vorticity $\omega(\rr)=0$ except at vortex cores where it is singular. However, as is well known from the study of Abrikosov vortex lattices, the coarse-grained velocity field \emph{can} acquire a rotational component as the system approaches a classical state~\cite{Fetter2001}, consistent with Bohr's correspondence principle.  Denoting the classical velocity field by $\vv_c(\rr)$, the rotational part can be described by the ansatz
\EQ{\label{vclass}
\vv_c(\rr)&=&\mathbf{\Omega}_c\times \rr,
}
for some $\bm{\Omega}_c=\Omega_c\hat{\mathbf{z}}$. Ignoring the vortex core structure,
the probability distribution for a cluster of radius $R_c$ exhibiting rigid-body rotation is found from \eref{Pv} as
\EQ{\label{vdep}
P(v)&=&\frac{n_0}{N_{\rm{tot}}}\int d^2\rr\; \delta(v-\Omega_c r)=\frac{2\pi n_0}{\Omega_cN_{\rm{tot}}}\int_0^{R_c}rdr \;\delta(r-v/\Omega_c)\nonumber\\
&=&\frac{2\pi n_0}{\Omega_c^2N_{\rm{tot}}} v\;H\left(\frac{\Omega_c R_c}{v}\right),
}
where the Heaviside function limits this behavior to the cluster interior. This distribution, with \eref{EHtildev}, gives the power-law form $E_H^{(0)}(k)\sim k^3$ for $k\leq k_c$, where
\EQ{\label{kcdef}
k_c=m\Omega_c R_c/\hbar,
}
giving a relation between the cluster size, classical vorticity field, and the $k^3$ scale 
range in the spectrum. Note that the $k^3$ form is quite distinct from the infrared result for a single vortex, for which \eref{Ppv} and \eref{EHtildev}, immediately yield $E_H^{(0)}(k)\propto k^{-1}$. 

For any finite system with a sufficiently high level of clustering, there will be significant modifications to the velocity field due to the interactions between the coherent structures and the boundary (or, equivalently, the image vortices that ensure the flow obeys the boundary conditions). For vortex distributions containing both positive and negative vortices, highly energetic, maximum-entropy configurations take the form of a macroscopic dipole \cite{Eyi2006.RMP78.87,Billam2013a}. The interaction between the clusters that form this dipole with their image vortices will induce a quadrupole mode in the phase, which will generate a stagnation point in the velocity field. The phase in the vicinity of a stagnation point may be modelled as $\theta(\rr) = \alpha x y$, where $\alpha$ is a constant with dimensions of inverse area~\cite{Recati2001a} (see Sec.~ \ref{sec:NumericalSampling} for examples of this phase structure). The velocity field is thus given by 
\begin{equation}
\mathbf{v}_s = \frac{\hbar\alpha}{m}(y,x),
\end{equation}
and from Eq. \eref{Pv}, one finds
\begin{equation}
P(v) = \frac{2\pi n_0 m^2}{\hbar^2 |\alpha|^2 N_{\rm tot}} v \; H\left(\frac{R_s \hbar |\alpha|}{mv}\right)
\label{PStagnation}
\end{equation}
with the Heaviside function limiting the behavior to some region $r < R_s$. Here $P(v)$ again yields the spectrum $E^{(0)}_H(k) \sim k^3$.  Eqs. \eref{vdep} and \eref{PStagnation} suggest that a significant region of the quantum kinetic energy spectrum may exhibit a $k^3$ power law if coherent vortex structures are present in the system. We also note from Eq.~\eref{Ekquant} that a $k^3$ spectrum corresponds to a constant momentum distribution. We explore the quantum kinetic energy spectrum further in the following section, where we examine vortex distributions via numerical sampling.

\section{Numerical sampling and spectral analysis of coherent vortex structures}
\label{sec:NumericalSampling}
In order to characterize the emergence of coherent vortex structures, we consider the end-states of freely decaying (i.e., unforced) 2DQT. Such states can be sampled using microcanonical methods~\cite{Billam2013a} and, in the case of appropriate experimental small-scale forcing~\cite{reeves_etal_prl_2013}, are expected to form via the subsequent freely-evolving vortex dynamics after switching off the forcing mechanism.


\begin{figure*}[!ht]
\includegraphics[width  = \textwidth,trim = 3.2cm 12.5cm 4.84cm 10.64cm, clip=true]{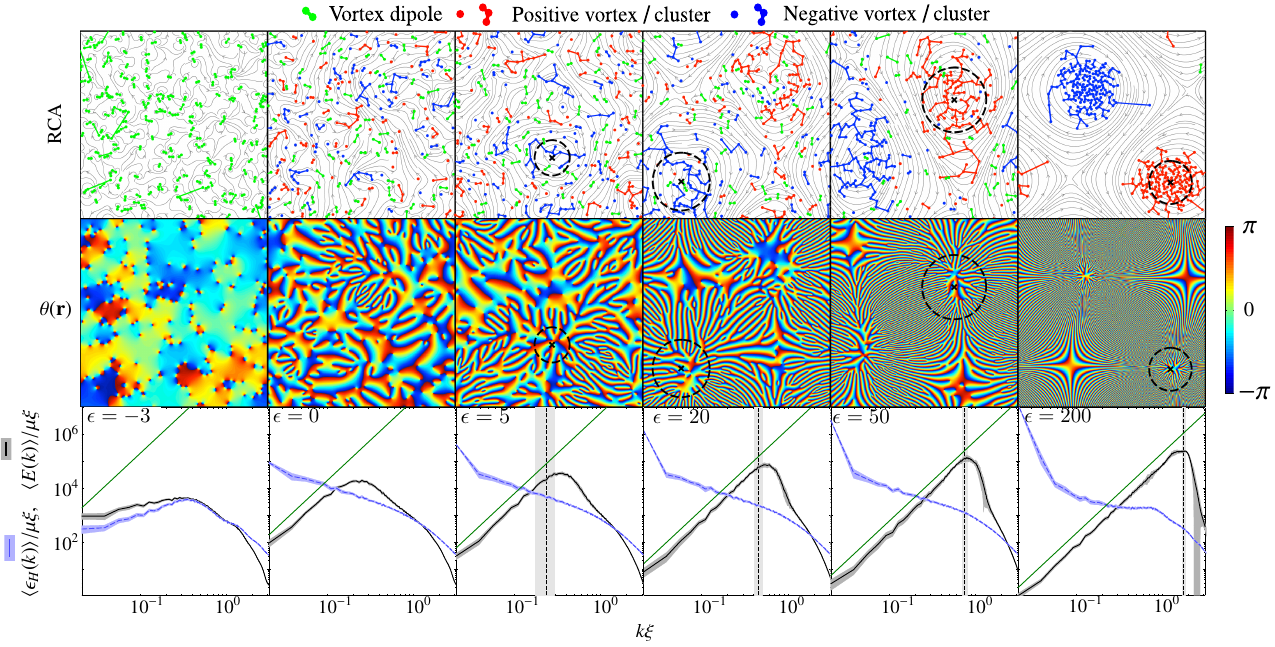}
\caption{ (Top row) Vortex distributions for a range of point-vortex energies. The distribution has been decomposed into clusters, dipoles and free vortices using the RCA~\cite{reeves_etal_prl_2013} (see text). For $\epsilon \geq 5$, $\mathbf{R}$ and $|\rr - \mathbf{R}| = R_c$ are shown by crosses and dashed lines respectively. Vortices in clusters are connected by solid lines showing the minimal spanning tree of the cluster. In each panel the field of view is $L\times L$, where $L=512\xi$. (Middle row) Phase profiles $\theta(\rr)$ for the corresponding vortex distributions. (Bottom row) Log-log graphs of the quantum kinetic energy spectrum $\langle E(k) \rangle$ [ensemble average of Eq.~\eref{Ekquant}], and the classical hydrodynamic spectrum $\langle\epsilon_H(k) \rangle$ [ensemble average of Eq.~\eref{ikeks}] for a range of point-vortex energies, averaged over $40$ random walk trajectories at  each energy. Shaded regions show $\pm1$ standard deviations. Lines proportional to $k^3$ (green) are shown for comparison. For $\epsilon\geq 5$ the vertical dashed line and shaded regions show $k^{\rm RCA}_c$, and $\pm1$ standard deviation respectively (see text).}
\label{SpectraandVortexDistribution}
\end{figure*}

\subsection{Microcanonical Sampling}
\label{MS}
 We consider neutral (zero net circulation) vortex configurations within a doubly-periodic box. We consider configurations of varying point-vortex energy per vortex, which correspond to varying degrees of clustering, and directly correspond to Gross-Pitaevskii energies in the incompressible regime \cite{Billam2013a}. Following Refs.~\cite{weiss_mcwilliams_pf_1991,Billam2013a}, the energy per vortex for a neutral configuration of $N$ point-vortices with charges $\{\kappa_j\} = \pm 1 $ (circulations $h\kappa_j/m$), located at positions $\{\rr_j\}$ within a square box of side length $L $ is given in terms of the vortex-pair energy
\begin{equation}
h(\rr) = \sum_{m = -\infty}^{\infty} \ln\left[\frac{\cosh(x-2\pi m) - \cos(y)}{\cosh(2\pi m)} \right] -\frac{x^2}{2\pi}
\label{eq:VortexPairEnergy}
\end{equation}
as
 \begin{equation}
\epsilon(\{\rr_j\},\{\kappa_j\}) = \epsilon_0 + \frac{1}{N}\sum_{p=1}^{N-1}\sum_{q=p+1}^N \kappa_p \kappa_q\; h\left(\frac{2\pi(\rr_p - \rr_q)}{L}\right),
\label{PVE}
\end{equation}
where the point-vortex energy is in units of $\Omega_0\xi^2$, $\Omega_0 = \pi\hbar^2n_0/m\xi^2$ is the unit of enstrophy in the homogeneous GPE, and $\epsilon_0 = -0.1170...$ is a constant that shifts the energy axis such that $\epsilon =0$ corresponds to an uncorrelated vortex distribution \footnote{This shifting of the energy axis is in fact equivalent to using the alternative expression for the point-vortex energy derived in~\cite{campbell_oneil_jsp_1991}, although we use Eq. \eref{PVE} for our computations as it is more convenient to work with numerically.}. 

The canonical momenta of the point-vortex system are the vortex coordinates (up to circulation prefactors) and, as a result, if the spatial domain is bounded so is the volume of accessible phase space~\cite{Onsager1949}. Consequently, in this system at $\epsilon = \epsilon_{\rm{max}} \approx - 0.255$ \cite{campbell_oneil_jsp_1991,Billam2013a} the structure function $W(\epsilon)$ (i.e.  the number  of available states at a given energy) reaches a maximum and  is monotonically decreasing thereafter as $\epsilon \rightarrow \infty$. Hence for $\epsilon > \epsilon_{\rm{max}}$, the temperature $T = W(\partial W /\partial \epsilon)^{-1}$  is formally negative. These negative temperature states are associated with a tendency for like-sign vortices to aggregate, and the emergence of  macroscopic vortex clusters in maximal entropy (equilibrium) configurations \cite{Eyi2006.RMP78.87}. Averaging over the microcanonical ensemble  at a  given energy (provided $N$ is sufficiently large, to ensure ergodicity) characterizes the end states of decaying 2DQT at that energy. 

The validity of this statistical description is dependent on the size (relative to the healing length), the point-vortex energy, and the vortex density of the system.
In particular, increasing the energy or vortex density, or decreasing the system size, eventually leads to strong coupling between vortex and sound degrees of freedom.
Note, however, that making the incompressible velocity everywhere small compared to the sound speed $c$ greatly reduces the strength of this coupling: a regime where the statistical description is valid can always be reached by increasing the system size (or alternatively reducing the vortex number, although this approach will eventually affect the system dynamics). Previous work has confirmed that the statistical approach correctly describes the end states of decaying turbulence in the damped GPE for energies up to $\epsilon=6$~\cite{Billam2013a} (in a box of length $512\xi$ for $384$ vortices). In this work we provide further confirmation of the approach for even higher energies in trapped systems (see Sec.~\ref{sec:Trapped}).

We investigate the equilibrium states over a range of point-vortex energies via a random-walk procedure. We start with an uncorrelated  distribution of $N=386$ vortices in a box of length $L=512\xi$. For an uncorrelated distribution the nearest-neighbour correlation functions $c_B = \sum_{p=1}^N \sum_{q=1}^B \kappa_p \kappa_p^{(q)} /BN$ (where $\kappa_p^{(q)}$ is the $q$th nearest neighbour to vortex $p$) are equal to zero~\cite{White2012a,reeves_etal_prl_2013,Billam2013a}. The vortices undergo a random walk towards a state with a desired point-vortex energy $\epsilon$, specified within a tolerance of $\Delta\epsilon = \pm0.01$. A minimum inter-vortex separation of $2\pi\xi$ is enforced to ensure the vortex cores are non-overlapping \cite{Bradley2012a}, such that higher order density gradients may be neglected, consistent with the analysis of Sec. \ref{sec:hydro}. This effective hard vortex core can also be viewed physically as an approximation to the energy barrier associated with the formation of multiply-quantized vortices in Gross--Pitaevskii theory \cite{Fetter2001}. 

Once a configuration with the desired energy is obtained, we find the corresponding Gross-Pitaevskii wavefunction using the constructive approach developed in Ref. \cite{Billam2013a}. In brief, the modulus of the wavefunction $\sqrt{\rho(\rr)}$ is constructed as the product of the individual vortex core wavefunctions [the numerical solution to Eq. \eref{yeq}]\, and the phase $\theta(\rr)$ is constructed from the sum of the phases due to individual vortex dipoles. We work in units of the healing length $\xi$ and the chemical potential $\mu$.  The wavefunctions are constructed on a numerical grid  with resolution  $M = 2048^2$. We sample within the range of point-vortex energies $\epsilon = [-3,200]$, sampling over 40 random walk trajectories at each value of $\epsilon$. 

\subsection{Recursive Cluster Algorithm}
To analyse our vortex distributions, we make use of the recursive clustering algorithm (RCA) presented in \cite{reeves_etal_prl_2013}. The algorithm yields detailed spatial information about  a particular vortex configuration by decomposing it into clusters, dipoles and free vortices. 
%
From the  decomposition of the vortex distribution, we may acquire characteristic information about each cluster. In particular, the physical location of the cluster is estimated by the center of mass 
\begin{equation}
\mathbf{R} = \frac{1}{N_c}\sum_{j \in C} \rr_j
\end{equation}
where $C$ denotes the set of all vortices that belong to a particular cluster, and $N_c \equiv |\kappa_c|$ denotes the number of vortices in the cluster. Additionally, the spatial extent of a cluster can be estimated by the cluster radius, which we define as the mean distance from the center of mass 
\begin{equation}
R_c = \frac{1}{N_c}\sum_{j \in C} |\rr_j - \mathbf{R}|.
\end{equation}
Although the cluster algorithm yields values for $\mathbf{R}$ and $R_c$ for every cluster in a given distribution, throughout this section we are primarily concerned with the \emph{largest} clusters in the system. Hereafter we will use the above notation to refer to the position and radius of the largest cluster only.

We show the resulting decomposition of the vortex distribution for a range of point-vortex energies in the top row of Fig.~\ref{SpectraandVortexDistribution}. Qualitatively, the algorithm captures the well-known physics of the point-vortex model:
at negative point-vortex energy, the distribution takes the form of a dipole gas,  with many or all vortices being bound in vortex-antivortex pairs. As the point-vortex energy is increased, clusters of same-sign vortices emerge, accumulating more vortices with increasing energy. With further increase of the point-vortex energy,  clusters continue to accumulate vortices, but also contract spatially,  storing more energy internally rather than accumulating more vortices from the remaining vortex field (as the latter would lower the entropy). At sufficiently high energy ($\epsilon \sim 200$) the phenomenon of supercondensation occurs \cite{Kra1980.RPP5.547}, and the distribution collapses into two macroscopic clusters, each of charge $N/2$. While such a state is unlikely to be achievable in atomic BEC, the supercondensed state nonetheless demonstrates the limiting physics  at very high energy.

\subsection{Spectral Analysis}
 In the bottom row of  Fig. \ref{SpectraandVortexDistribution} we show the (ensemble-averaged) quantum kinetic energy spectrum $\langle E(k) \rangle$ [Eq. \eref{Ekqdef}] over a range of point-vortex energies. It is evident that, for positive energies, the spectrum acquires a $k^3$ scaling in the infrared. The scaling begins at low $k$ and progresses towards larger wavenumbers  as the point-vortex energy of the system is increased. The behavior of the spectrum is in stark contrast with the classical spectrum [Eq. \eref{ikeks}], for which the emergence of large-scale structure is signified by a spectral ``pile-up" at length scales of order the system size, as is shown in Fig.~\ref{SpectraandVortexDistribution}. Notice however that the two spectra are very similar in the low energy vortex-dipole regime. Even for the relatively modest point-vortex energy $\epsilon = 5$ the spectrum exhibits nearly a decade of $k^3$ scaling in this system. For sufficiently high energy ($\epsilon \approx 100$) the range of the $k^3$ scaling extends past the point $k=\xi^{-1}$. In \emph{dynamical} simulations, it is unlikely that the $k^3$ scaling would extend this far, as effects due to compressibility are non-negligible at velocities comparable to $c$ (see Sec.~\ref{sec:Trapped}). At high energy, the regions of slowly-varying phase that contribute to the infrared spectrum are clearly  seen at the stagnation points and in the interior regions of the largest clusters. In the supercondensed state, the stagnation point  phase structure $\theta(\rr) = \alpha x y$    becomes particularly evident.
\begin{figure}[!t]
\includegraphics[width = \columnwidth]{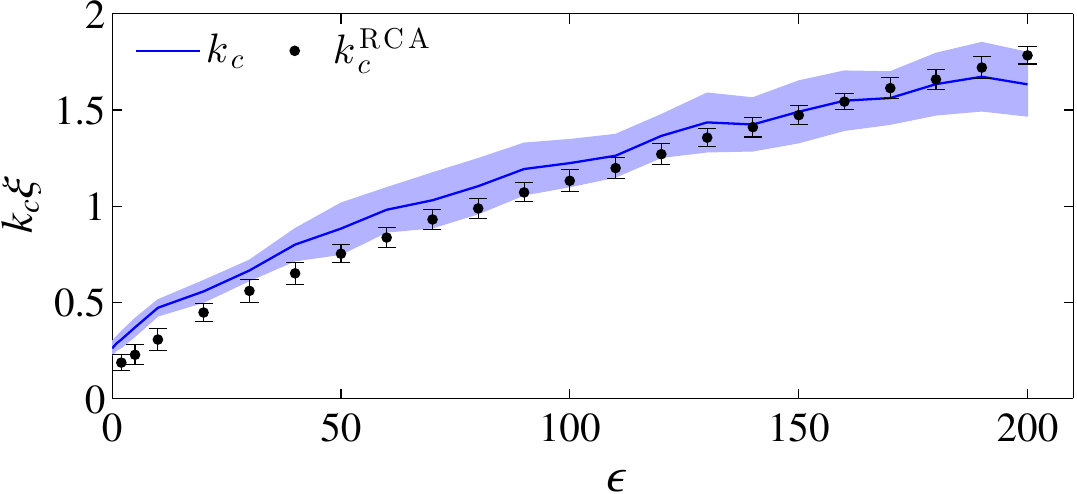}
\caption{Wavenumber corresponding to the peak of the spectrum $k_c$, as a function of the point-vortex energy per vortex $\epsilon$. The shaded region and errorbars show $\pm$ 1 standard deviations.}
\label{PeakvsE}
\end{figure}

The location of the peak in the kinetic energy spectrum, which we label $k_c$, gives an indication of the $k^3$ scale-range observed in the negative-temperature regime. In light of Eq. \eref{EHtildev}, in the hydrodynamic approximation where $v=\hbar k/m$,  this wavenumber indicates a  most probable or ``characteristic" velocity. This characteristic velocity is associated with the coherent structures; as the energy is increased and large structures emerge, we observe that phase fluctuations of a characteristic wavelength develop around these structures, and  throughout the system [Fig.~\ref{SpectraandVortexDistribution}].  The characteristic wavelength of the fluctuations shortens as $\epsilon$ increases. These observations are consistent with Onsager's prediction -- that the velocity field of a negative-temperature equilibrium state will be dominated by the coherent motion of the macroscopic vortex clusters, since the motion of the remaining vortices is essentially random \cite{Eyi2006.RMP78.87}.

We find that the characteristic velocity can be estimated from the RCA  by considering only the largest cluster (in terms of charge) in a given configuration.
The wavelength of the phase variations at a distance $r$ from a large cluster will be $\lambda = 2\pi r/  \kappa_{\rm{net}} $, where $\kappa_{\rm net}$ denotes the net charge within the region enclosed by a circle of radius $r$. The corresponding wavenumber is thus  $k  = \kappa_{\rm{net}} / r$.

  To calculate a value for $k_c$ from the RCA data, which we denote $k_c^{\rm RCA}$, we take the largest cluster and calculate the net charge enclosed within the cluster's radius $R_c$. We only consider energies at which we may unambiguously define the largest cluster ($\epsilon \geq 5$ \footnote{We analyze  energies for which there is a unique $N_c = \max(\{N_{c,i}\})$ for at least $90\%$ of the random walk trajectories. In the few cases where multiple clusters satisfy $N_c = \max(\{N_{c,i}\})$, one of these clusters is selected at random. }); for these values, $k^{\rm RCA}_c$ is shown in Fig.~\ref{SpectraandVortexDistribution} (bottom row), where it clearly provides a good indication of the location of the spectral peak. Conversely, the location of the peak provides a good estimator of the scale of the largest cluster in the system. We also compare $k_c$ as calculated directly from the kinetic energy spectrum to  $k_c^{\rm RCA}$ in Fig.~\ref{PeakvsE}. In general there is very good agreement in the data, even though the RCA does not account for cluster anisotropy [e.g. see Fig.~\ref{SpectraandVortexDistribution}, $\epsilon = 50$]. Agreement is poorest at lower energies ($\epsilon \lesssim 10$), when the largest clusters are not significantly larger than those in the background vortex distribution. However, as the energy is increased, agreement improves as the largest clusters begin to dominate the velocity field.

\section{Emergence of rigid-body rotation}
\label{sec:ClassicalFlows}

\begin{figure}[!t]
\includegraphics[width  = \columnwidth]{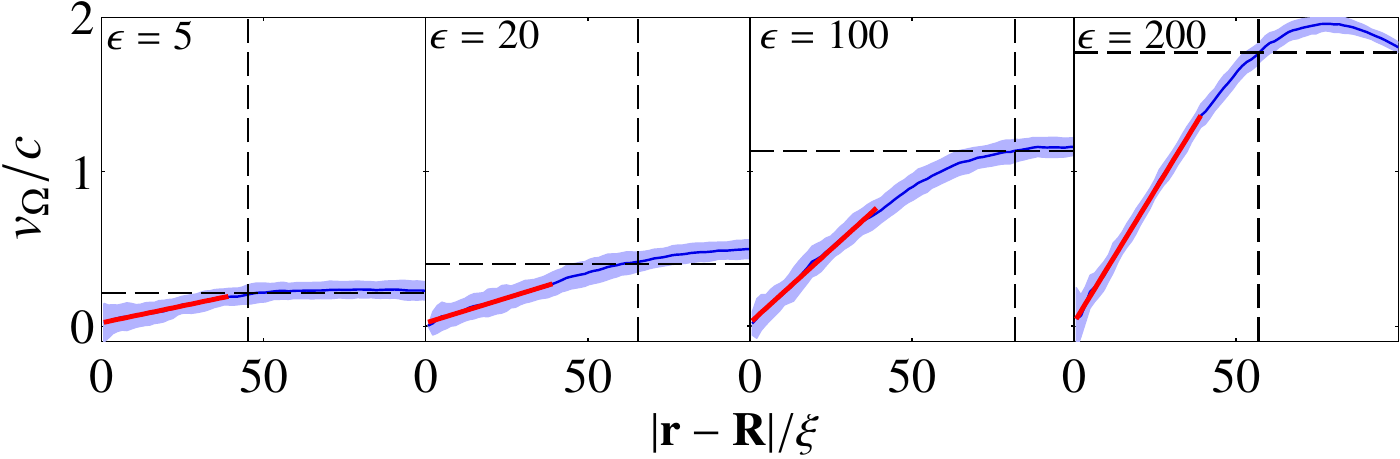}
\caption{ Azimuthal velocity field  $v_\Omega$ of the largest cluster, as a function of distance from the cluster center $|\rr-\mathbf{R}|$, for a range of point-vortex energies $\epsilon$. Shaded regions show $\pm1$ standard deviations. Solid (red) lines show a linear fit to the averaged velocity field within the region $ 0 \leq |\rr - \mathbf{R}|\leq 40\xi$. The slope yields a value for $\Omega_c^{\rm fit}$ (see text and Fig.\ref{OmegavsE}). Horizontal and vertical   dashed lines show the values $ v_c^{\rm RCA} /c  \equiv   k_c^{\rm RCA}\xi$ (see Fig.~\ref{PeakvsE}) and $\langle R_c \rangle$ respectively.  }
\label{Vphi}
\end{figure}

\subsection{Azimuthal velocity field}
According to the analysis in Sec. \ref{Signatures}, the presence of a $k^3$ spectrum suggests that the azimuthal velocity field in the vicinity of a large cluster  may mimic that of a rigid body, that is $v_\theta(\rr^\prime) = \Omega_c |\rr^\prime|$, where $|\rr^\prime| = |\rr - \mathbf{R}|$ is the distance from the cluster center. However,  we have shown that the $k^3$ spectrum may also be due to the stagnation points of the velocity field. Due to the non-local nature of the Fourier transform, it is difficult to disentangle the spectral contributions from rigid-body rotation and the stagnation points. This  motivates us to analyse the clusters directly, to determine the extent to which they exhibit rigid-body characteristics.

 Again considering energies $\epsilon \geq 5$, we calculate the angular velocity field relative to  the cluster center, averaging over the azimuthal direction and the ensemble:
\begin{equation}
v_\Omega(r^\prime)  = \left \langle{ \frac{1}{2\pi} \int_0^{2\pi} \; d\theta'\; |v_\theta(\rr^\prime) |} \right\rangle.
\end{equation}

The resulting velocity field for a range of point-vortex energies is shown in Fig.~\ref{Vphi}.  The averaged velocity field is approximately linear over the cluster interior, although fluctuations are larger at lower energy. The linear behaviour is typically maintained up to scales comparable to the average cluster radius $\langle R_c \rangle $. At $\langle R_c \rangle $ the velocity is well approximated by the characteristic velocity $v_c^{\rm RCA} \equiv\hbar k_c^{\rm RCA}/m$. We find $v_\Omega$ to be linear for at least $r^\prime\lesssim 40\xi$ for  $\epsilon\geq5$, and that the slope of $v_\Omega$ steepens with increasing $\epsilon$.   Note that although the RCA sometimes over estimates the region of linear behavior, the velocity is always well described by $v_c^{\rm RCA}$ at $\langle R_c\rangle$. Additionally, notice $v_\Omega(0) \approx 0$, suggesting that the RCA accurately determines the location of the cluster center. 

\subsection{Measures of classical vorticity}
\begin{figure}[!t]
\includegraphics[width = 1\columnwidth]{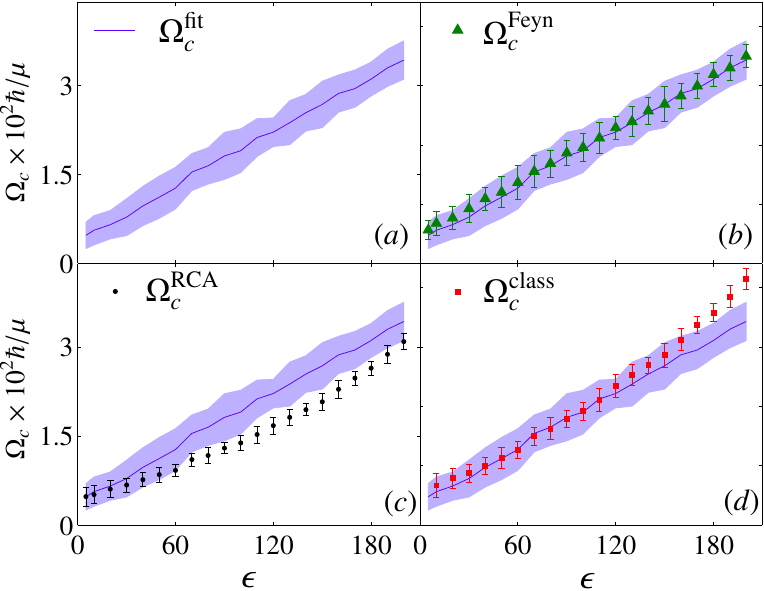}
\caption{Measures of the rigid-body rotation frequency $\Omega_c$. The shaded regions ($\Omega_c^{\rm fit}$) and error bars (other measures) indicate $\pm 1$ standard deviations. In (b)-(d) $\Omega_c^{\rm fit}$  is shown for comparison.}
\label{OmegavsE}
\end{figure}

To further characterize the rigid-body flow field, we may determine the rigid-body rotation frequency $\Omega_c$, thus also determining the characteristic turnover time $T = 2\pi/\Omega_c$ for the largest cluster.  We describe four measures:

$\Omega_c^{\rm{fit}}$: A value may be obtained from the slope of a linear fit to $v_\Omega$. We fit over the region $0 \leq r^\prime\leq 40\xi$, where linearity holds for all $\epsilon$, as shown by the lines of best fit  in Fig. \ref{Vphi}. We show $\Omega_c^{\rm{fit}}$ as a function of $\epsilon$ in Fig.~\ref{OmegavsE}(a). There is a clear linear trend in the data, which are well described by the relation $\Omega_c^{\rm{fit}} = (1.5\epsilon + 36) \times 10^{-4}$. We use $\Omega_c^{\rm fit}$ as a base measure, which we compare against other measures.

$\Omega_c^{\rm{Feyn}}$: The positive-temperature ground state for a system rotating at frequency $\Omega_c$ is an ordered vortex lattice that has a constant vortex density given by Feynman's rule, $n_v=m\Omega_c/\pi\hbar$ \cite{Feynman1955}. In order to maintain rigid-body rotation, the negative-temperature clustered states considered in this work must still exhibit a constant area per vortex on average, even though they do not maintain crystalline order. Applying Feynman's argument, we expect a rotation frequency
\EQ{\label{Varea}
\Omega_c^{\rm Feyn}=\frac{\pi\hbar \langle n_v\rangle }{m},
}
where $\langle n_v\rangle$ denotes the average vortex density of the largest cluster in each sample, averaged over the ensemble.
Considering the cumulative distribution $N_v(r)$, which counts the number of vortices with $|\rr_i - \mathbf{R}| \leq r$, 
 we find the distribution is well described by $N_v(r) = \pi n_v r^2$ (as required for constant $n_v$) over the region where $v_\Omega$ is linear. We verify that this is not an artifact of the vortex-separation minimum by reducing the separation cutoff used in our microcanonical sampling from $2\pi\xi$ to $\xi$, finding nearly identical results. Fitting over the same region $0\leq r^\prime \leq 40\xi$  yields values for $\Omega_c^{\rm Feyn}$ which are in excellent agreement with $\Omega_c^{\rm fit}$, as shown in Fig. \ref{OmegavsE}(b).  We emphasize that $\Omega_c^{\rm fit}$ is obtained from the velocity field, while $\Omega_c^{\rm Feyn}$ is determined by the vortex distribution.

$\Omega_c^{\rm{RCA}}$: A value for $\Omega_c$ may also be calculated from the RCA data. Since the rigid-body velocity field persists up to the characteristic wavenumber $k_c = m\Omega_c R_c/\hbar$, and $k_c$ is well described by $k_c^{\rm{RCA}} = \kappa_{\rm{net}} / R_c$, clearly we may consider \begin{equation}
\Omega_c^{\rm{RCA}} \equiv  \left \langle \frac{\hbar \kappa_{\rm{net}} }{m R_c^2} \right \rangle=\frac{1}{2} \left \langle \frac{1}{\pi R_c^2} \int_\mathcal{R} d^2\rr \; \omega(\rr) \right \rangle,
\end{equation}
where $\mathcal{R}$ is the region enclosed by a circle of radius $R_c$ centered on $\mathbf{R}$.
This  is equivalent to averaging the vorticity distribution over the region of the cluster.
The values $\Omega_c^{\rm{RCA}}$ are  shown in Fig.~\ref{OmegavsE}(c). There is reasonable quantitative agreement between the data obtained directly from the wavefunction ($\Omega_c^{\rm fit}$) and that extracted from the RCA. It is clear that there is some discrepancy in qualitative trend however, and agreement between the two quantities is poorest for $\epsilon \sim 120$. This ``sag" at intermediate energies is due to a decline in vortex density in the outer region of the cluster, which causes $N_v(r)$ to  deviate from the expected quadratic behavior, and consequently  $v_\Omega$ to deviate from rigid-body behavior (Fig.~\ref{Vphi}, $\epsilon = 100$). For perfect rigid body rotation extending out to $R_c$, one would expect $\Omega_c^{\rm fit}$ and $\Omega_c^{\rm RCA}$ to yield exactly the same values. 
 Thus $\Omega_c^{\rm RCA}$ indicates the extent to which the velocity field deviates from perfect rigid-body rotation over the scale of the cluster as defined by the RCA value $R_c$.
 \begin{figure*}[!t]
 \includegraphics[width = \textwidth, trim = 16.5cm 16cm 16.5cm 10cm, clip = true]{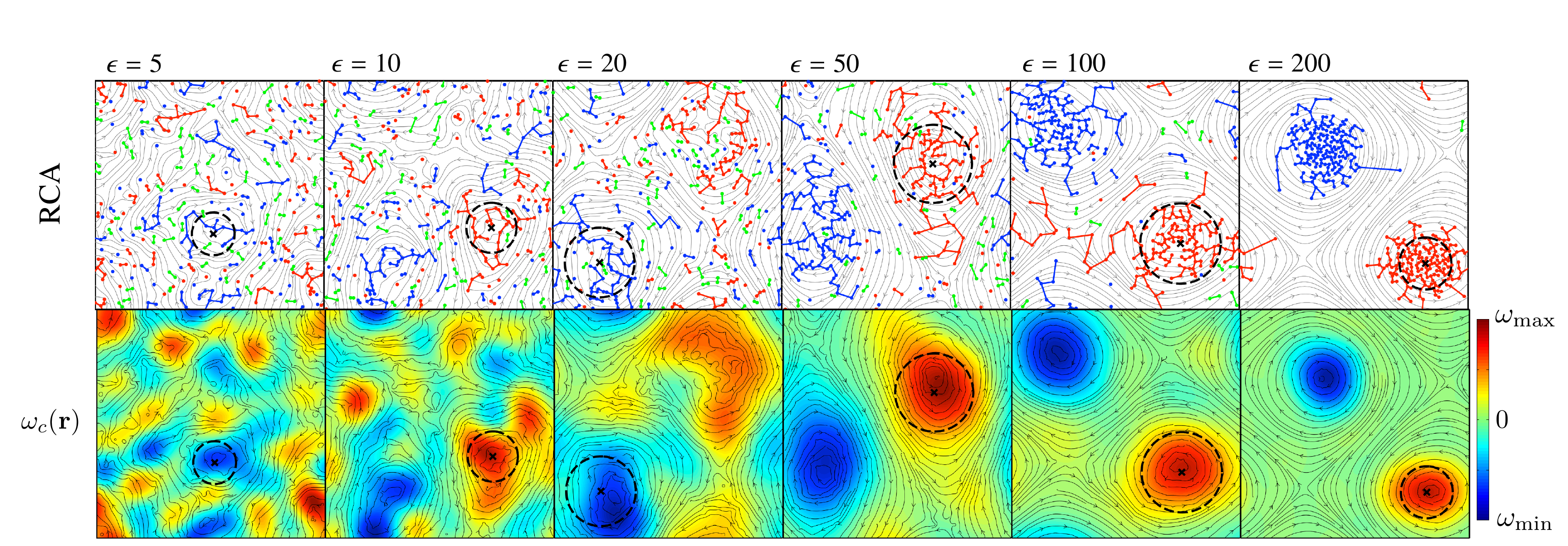}
 \caption{The emergence of classical flows: Exemplar RCA vortex distributions and  coarse grained vorticity fields $\omega_c(\rr)$ [the latter obtained by including only spatial modes $k \leq k_{\rm{cut}} = 2\pi/(2R_c)$] for a range of $\epsilon$. Streamlines show the velocity field $\vv(\rr) = \hbar \nabla \theta(\rr)/m$. Crosses show the center-of-mass position of the largest cluster in each sample, $\mathbf{R}$, as determine by the RCA.  The RCA radius $R_c$ is shown by the dashed line.}
 \label{fig6}
 \end{figure*}
 \par
$\Omega_c^{\rm{class}}$: The presence of  a rigid-body velocity field requires that, under an appropriate \emph{coarse graining} procedure (see also, e.g., \cite{Baggaley2012,Baggaley2012b}), the  formally singular vorticity field becomes constant over the central region of the cluster, i.e.,  $\omega(\rr) = h/m\sum_i \delta(\rr-\rr_i)  \rightarrow\omega_c(\rr) \simeq 2\Omega_c$ for $|\rr-\mathbf{R}| \lesssim R_c$.  Defining an average over many quantum vortices, we consider the coarse-grained \emph{classical} vorticity field
\begin{equation}\label{wc1}
\omega_c(\rr) \equiv \frac{1}{2\pi} \int_{\mathcal{K}} d^2\mathbf{k}\; \tilde{\omega}(\mathbf{k})  e^{i \mathbf{k} \cdot \rr}
\end{equation}
where $\tilde{\omega}(\kk) = \mathcal{F}[\omega(\rr)]$ and $\mathcal{K}$ is the $k$-space domain satisfying $|\mathbf{k}| < k_{\rm{cut}} = 2\pi/\ell_{\rm cut}$ for a chosen  cutoff length-scale $\ell_{\rm{cut}}\gg\xi$. Coarse graining over the spatial extent of the largest vortex cluster ($\ell_{\rm{cut}} = 2R_c$) yields a value consistent with the other measures if we consider $\omega_c(\mathbf{r})$ at the cluster center:
\begin{equation}
\Omega_c^{\rm{class}} = \left\langle \frac{1}{2}\omega_c(\mathbf{R}) \right\rangle.
\end{equation}
Values for $\Omega_c^{\rm class}$ are shown in Fig.~\ref{OmegavsE}(d).
We find excellent agreement between $\Omega_c^{\rm{class}}$ and $\Omega_c^{\rm{fit}}$ until $\epsilon \sim 150$, at which point $\Omega_c^{\rm{class}}$ deviates to higher values. This discrepancy is due to taking the value locally at $\mathbf{R}$, where the vorticity can be slightly more concentrated, particularly for large clusters, whereas $\Omega_c^{\rm{fit}}$ incorporates information away from the cluster center.  $\Omega_c^{\rm class}$ indicates that, apart from at very high energy, the clusters do not deviate significantly from rigid-body motion in their interior.

Coarse graining over $\ell_{\rm cut} = 2\bar{d}$ for mean nearest-neighbour distance $\bar{d}$ (typically  $\bar{d} \simeq R_c/3$) yields similar values for $\Omega_c^{\rm class}$, but with larger fluctuations. For this value of $\ell_{\rm cut}$ we also verify that $ \omega_c(\rr')  \approx \rm{const.}$ for $|\rr'| \lesssim R_c$. Averaging over the azimuthal direction and ensemble to obtain $\omega_c(r^\prime)$, we find $\langle \omega_c(40\xi)\rangle \gtrsim 0.9\langle \omega_c(0)\rangle$ at all energies, consistent with the linear $v_\Omega$ and quadratic $N_v$ observed up to this scale.  At the cluster radius,  $\langle \omega_c(R_c)\rangle/ \langle \omega_c(0)\rangle$  $\sim 0.6-0.8$ for low ($\epsilon \lesssim 50$), and high ($\epsilon \gtrsim 150$) energies respectively, and $\langle \omega_c(R_c)\rangle/ \langle \omega_c(0)\rangle \sim 0.5$ for intermediate energies. This is consistent with the deviation from rigid-body behavior at larger scales as seen in Fig.~\ref{Vphi}, and as indicated by the qualitative trend of $\Omega_c^{\rm RCA}$.

The emergence of a classical flow field from the quantum vortex distribution is qualitatively captured in Fig. \ref{fig6}, where we present particular examples of the vortex distribution (as decomposed by the RCA) and the classical vorticity field $\omega_c(\rr)$ (for $\ell_{\rm cut} = 2R_c$), generated at various $\epsilon$. As large  clusters emerge, they generate macroscopic regions of approximately uniform vorticity, capturing the qualitative features of the emergent quasi-classical velocity field,  as shown by the velocity streamlines. 


It is interesting to compare the rigid-body rotation property observed here to the rotational properties of coherent structures in classical fluids. The states we have considered here are the freely-decayed states of a turbulent superfluid. Decaying turbulence described by the inviscid Euler equations has been shown to approach the statistical equilibrium predicted by the mean-field Montgomery-Joyce (sinh-Poisson) equation \cite{montgomery_etal_pf_1992,Montgomery1974a}. With quantum vortices we would expect to recover this regime  in the limit $N\rightarrow \infty$. Indeed, the rigid-body rotation observed here in quantum vortex clusters for large $N$ is qualitatively consistent with the form of the doubly-periodic vortex dipole solution of the Montgomery-Joyce equation~\cite{Pasmanter1994a}.

\section{Dynamical Emergence in a Trapped System}
\label{sec:Trapped}

Finally, we demonstrate that coherent structures and the associated $k^3$ power law can emerge dynamically in a trapped system, and compare the dynamical results against sampling. The model we use to describe the dynamics of a 2D Bose gas is the damped Gross-Pitaevskii equation (dGPE):
\begin{equation}
i\hbar\frac{\partial \psi(\brr,t)}{\partial t}=(1-i\gamma)(\mathcal{L}-\mu)\psi(\brr,t),
\end{equation}
where 
\begin{equation}
\mathcal{L} =  \left(-\frac{\hbar^2\nabla_\perp^2}{2m}+V_{\rm{ext}}(\rr) + g_{2}|\psi(\brr,t)|^2\right),
\end{equation}
for an external confining potential $V_{\rm ext}(\rr)$. The damping parameter
$\gamma$ describes the finite-temperature effects due to collisions between the
condensate and a stationary thermal reservoir with chemical potential $\mu$.
Within the framework of $c$-field theory, the dGPE can be obtained from the
stochastic projected Gross-Pitaevskii equation for a large BEC in the low-temperature regime, for which the thermal noise is negligible~\cite{Blakie08a}.  The dGPE has been used extensively in
quantum turbulence studies \cite{Kobayashi05a, Neely10a, Numasato10a, Reeves12a, neely_etal_prl_2013}, and has been demonstrated to give
qualitative agreement with experiment even for relatively high
temperatures~\cite{Neely10a}. We set $\gamma = 10^{-5}$, an experimentally
realistic value under the conditions for which the dGPE theory is
valid~\cite{Bradley2012a}. For such a value of $\gamma$, the modifications to
the vortex dynamics \cite{tornkvist_shroder_prl_1997} are essentially
negligible over the integration time we consider. The primary effect of the
damping is thus to suppress compressible excitations at very high $k$, which
are numerically demanding to resolve, yet have little physical effect on the vortex dynamics in the regime of interest here.

We simulate a 2D BEC confined within a circular well or ``bucket'' potential of radius $R$, i.e.,
\begin{equation}
\label{BucketTrap}
V_{\rm ext}(\rr) = V_0\{1 +  \tanh[(r - R)/a]\},
\end{equation}
and set $V_0/\mu=10$, $R/\xi=200$, and $a/\xi=3$, such that $V_{\rm ext}$ approximates a hard-wall potential. We remark that Bose condensation in a quasi-uniform cylinder has been recently demonstrated experimentally \cite{Gaunt2013a}.

In this system,  the energy of an $N$-vortex configuration may be characterized by the energy per vortex for point-vortices (in units of $\Omega_0\xi^2$) within a circular domain $D$ of radius $R$ \cite{yatsuyanagi_etal_prl_2005}:
\begin{multline}
\label{PVEcircle}
\epsilon_{\circ}(\{\rr_j\},\{\kappa_j\}) = \tilde{\epsilon} -\frac{1}{N}\sum_{p=1}^{N}\sum_{q\neq p}^{N} \kappa_p\kappa_q \ln \left|\frac{\rr_p-\rr_q}{\xi}\right| \\ - \frac{1}{N} \sum_{p=1}^N\sum_{q=1}^N \kappa_p\bar{\kappa}_q \ln \left|\frac{\rr_p-\bar{\rr}_q }{\xi}\right|,
\end{multline}
where $\bar{\rr}_j = R^2\rr_j/|\rr_j|^2$ is the location of an image vortex with charge $\bar{\kappa}_j = -\kappa_j$ and $\tilde{\epsilon} \approx -4.158$ shifts the axis such that $\epsilon_\circ = 0$ corresponds to an uncorrelated distribution, as does $\epsilon_0$ in Eq. \eref{PVE}. 
The images ensure that the velocity field satisfies the boundary condition $\vv \cdot \hat{\mathbf{n}} \;|_{\partial D} = 0$, i.e., that the flow perpendicular to the boundary $\partial D$ is zero everywhere on $\partial D$.

We set the vortex number to $N = 112$. Whilst fewer vortices will inevitably lead to larger statistical fluctuations, a lower vortex density is beneficial here as it reduces the radiative loss of vortex energy into the sound field. Reducing the vortex density ensures that the incompressible velocity field is everywhere small compared to the speed of sound $c$,  and also lowers the chance of vortex core overlap, thus ensuring that the vortices remain sufficiently well separated such that their Gross--Pitaevskii dynamics are well-approximated by a point-vortex description \cite{Fetter1966}. 



\begin{figure}
\includegraphics[width = \columnwidth]{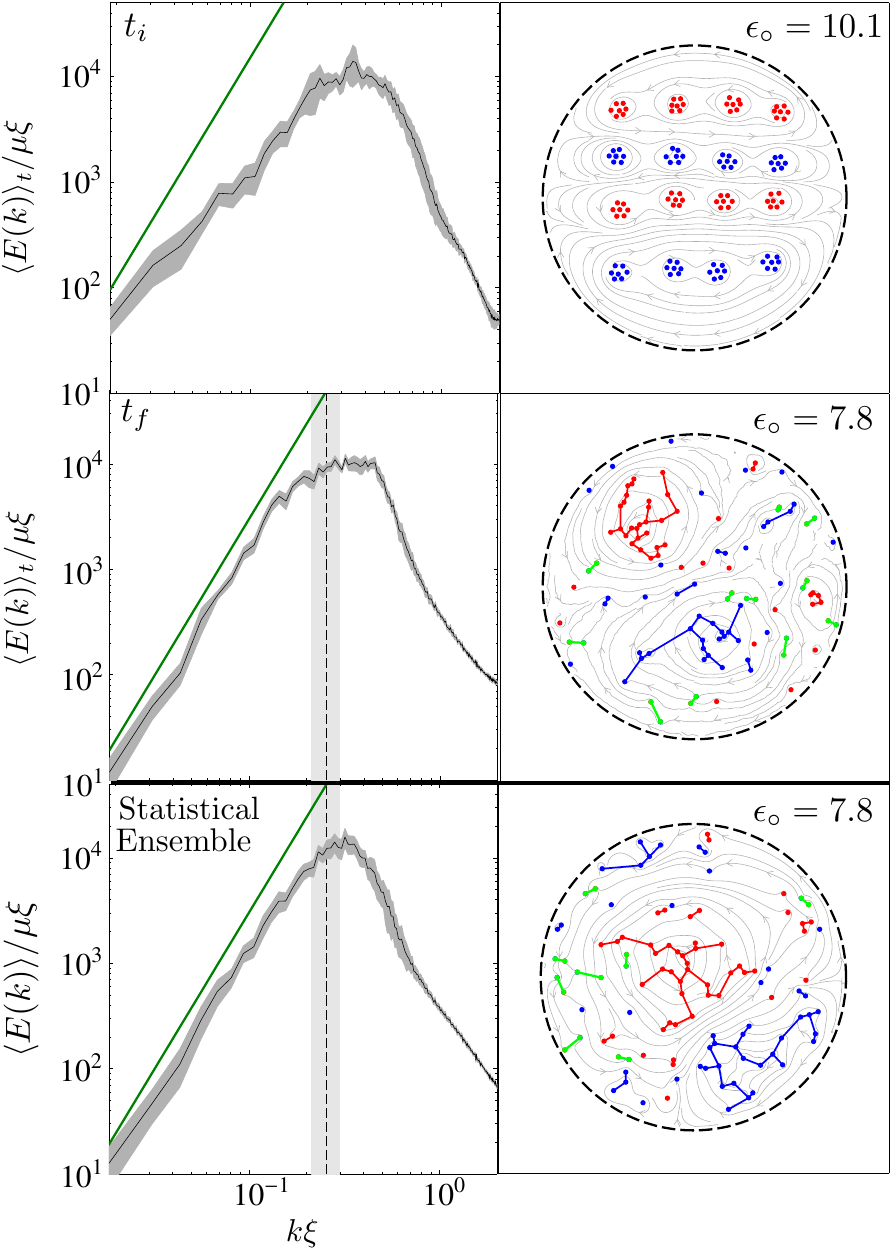}
\caption{ Time-averaged quantum kinetic energy spectra (left) and vortex distributions (right) for the dynamical system at $t_i=0\hbar/\mu$ (top row) and $t_f = 11300\hbar/\mu$ (middle row). The spectra for the dynamical system are averaged from $t$ to $t+500\hbar/\mu$. The spectrum and distribution at $t_f$ can be compared against those of the statistical ensemble (40 samples) at the same point-vortex energy (bottom row).  On the spectrum plots, lines proportional to  $k^3$ are shown for comparison, shaded regions show $\pm1$ standard deviation, the vertical dashed line shows $k_c$ as calculated from the ensemble using Eq.~\ref{kcdef}, with the shaded vertical band showing $\pm1$ standard deviation. Symbols for the vortex distribution plots are as denoted in Fig.\ref{SpectraandVortexDistribution}}

\label{dynamics}
\end{figure}

The initial condition, $\psi_\circ(\rr,0)$, is generated via a similar method to that outlined in Sec. \ref{sec:NumericalSampling}:
\begin{equation}
\psi_\circ(\rr) = \psi_{\rm TF}(\rr) e^{i\theta_\circ(\rr)} \prod_{j= 1}^N \chi(|\rr -\rr_j|/\xi).
\end{equation}
Here the Thomas-Fermi wavefunction $\psi_{\rm TF}(\rr)=\sqrt{(\mu -V_{\rm ext}(\brr))/g_2}$ for $V_{\rm ext}(\brr)\leq\mu$ and $0$ otherwise, $\chi(r)$ is the radial core profile of an individual quantum vortex [the numerical solution to Eq. \eref{yeq}], and
\begin{equation}
\theta_\circ(\rr) = \sum_{j=1}^N \kappa_j\theta_j(\rr) + \bar{\kappa}_j{\theta}_j(\bar{\rr}),
\end{equation}
where $\theta_j(\rr)$ is the phase due to an individual positive vortex at $\rr_j$.  We prepare a high-energy [$\epsilon_\circ(t_i) = 10.1$], non-equilibrium initial vortex configuration, as shown in Fig~\ref{dynamics}. The wavefunction is constructed on a spatial domain of length $L=512\xi$, and is discretized on  a grid of $M = 1024^2$. We numerically integrate the dGPE pseudo-spectrally, using an adaptive 4th-5th order Runge-Kutta method \cite{NumericalRecipes,XMDS}, and a relative error tolerance of $\tau=10^{-5}$. We have confirmed that all statistical measures of the vortex dynamics remain the same for $L=700$, $M=2048$, $\tau=10^{-6}$.

 Full dynamical evolution of the condensate density and phase, RCA vortex distribution, and quantum kinetic energy spectrum are provided in the supplemental material~\cite{Note5}. The initial configuration is highly unstable, and  the horizontal lines of like-charge clusters undergo a \emph{roll-up}, due to the Kelvin-Helmholtz instability. The initial kinetic energy spectrum, shown in Fig. \ref{dynamics}, does not follow the $k^3$ power law, but does exhibit a well-defined peak. The peak is due to the initial clustering in the configuration, which produces phase fluctuations of a characteristic wavelength $\lambda \sim 10\xi-20\xi$ \cite{Note5}.

Full relaxation to equilibrium is slow, requiring an integration time of order $10^4\hbar/\mu$. However, clear signs of a $k^3$ power law emerge after $t\sim3000\hbar/\mu$, as, at this stage of the evolution, large, quasi-equilibrium clusters have already formed~\cite{Note5}. This is consistent with the observations in Ref.~\cite{Billam2013a}, where it is shown that the macroscopic clusters begin to form well before the equilibration time. After $t_f=11300\hbar/\mu$ of evolution, the system has clearly reached the equilibrium state containing coherent vortex structures, as shown in Fig \ref{dynamics}. Although during the evolution some of the vortex energy is lost to sound, most of the vortex energy is retained  [$\epsilon_\circ(t_f) \sim 7.8$], and thus the configuration is still well within the negative temperature regime. It may be that radiative energy loss of the vortex distribution as a whole is partially inhibited by vortex ``cross-talk", a mechanism via which vortices of the same sign can efficiently impart energy to each other through radiation and absorption of sound~\cite{Parker2012}. 

We emphasize that the final energy per vortex $\epsilon_\circ =7.8$ is substantial, being analogous to a point-vortex energy in the periodic system considered in Sec. \ref{sec:NumericalSampling} of $\epsilon \gg 7.8$. Indeed, using the supercondensation energy ($\epsilon_\circ \approx 70$) as a reference, one can consider the final point-vortex energy in the simulation, $\epsilon_\circ(t_f) = 7.8$,  to be roughly equivalent (interpreted as a fraction of the supercondensation energy) to $\epsilon \approx 22$ in the doubly-periodic system considered in Sec.~\ref{sec:NumericalSampling} (where supercondensation occurs at $\epsilon \approx 200$). A more quantitative estimate of equivalence can be obtained using the nearest neighbour clustering measure $c_1$ (as defined in Sec.~\ref{MS}); we find the equilibrium values to be approximately the same ($c_1 = 0.35$) for $\epsilon_\circ = 7.8$ and $\epsilon \approx 25$, supporting the above analysis.

The quantum kinetic energy spectrum and vortex distribution at $t_f$ can be compared against the same quantities obtained from a statistical ensemble, as shown in Fig.~\ref{dynamics}. The spectra are qualitatively very similar, and nearly identical in the $k^3$ region. The RCA value for $k_c$ still gives a reasonable indication of the range of $k^3$ behaviour and the location of the spectral peak. The vortex distribution as determined by the RCA is also very similar [see Fig.~\ref{dynamics}].

We propose that the most direct way initial conditions similar
to those shown in Fig.~\ref{dynamics} could be created is by carefully-controlled laser stirring and manipulation
protocols: The field of two-dimensional quantum turbulence has seen several
numerical studies of the injection of clustered vortices via optical stirring
potentials in recent years \cite{White2012a, WhiteTopological2013, Reeves12a,
White2013a, Wilson2013, Stagg2014}, and injection of small vortex
clusters has indeed already been demonstrated experimentally~\cite{Neely10a}.
While neutral systems having similar spatial extent (relative to the healing length) and containing as many vortices as we have
considered here may be challenging to achieve, they are nonetheless
within the scope of current experimental technology.

 \section{Conclusion}
\label{sec:Conclusion} 
In this work we have shown that the coherent vortex structures that emerge in decaying 2DQT can exhibit quasi-classical rigid-body rotation, obeying the Feynman rule of constant areal vortex density while remaining spatially disordered.  By developing a rigorous link between the velocity probability distribution and the quantum kinetic energy spectrum we have shown that these coherent structures are associated with a $k^3$ power-law in the infrared region of the spectrum. The power-law region terminates at a peak located the inverse spatial scale of the largest cluster. The $k^3$ spectrum and associated peak  provide signatures of coherent structure formation that may be measured independently of the vortex configuration data, and should be accessible in atomic BEC experiments. Furthermore, our analysis illuminates the important distinction between the quantum kinetic energy spectrum and the power spectrum of the velocity autocorrelation function in a quantum fluid. 

Although individual vortex cores can be observed, charge-sensitive vortex detection remains a significant challenge in atomic BEC experiments.
By identifying a clear spectral signature that may be accessible in ballistic expansion imaging, our work paves the way for experimental observations of negative-temperature coherent vortex structures in two-dimensional atomic Bose-Einstein condensates. Experimental confirmation of the Feynman rule at negative temperature may provide further indication of the appearance of rigid-body rotation and the universality of rotational velocity fields in quantum turbulence.



\section*{Acknowledgements} 
We thank Sam Rooney and Hayder Salman for comments on the manuscript, and Gavin King for useful discussions. We are supported by the Marsden Fund of New Zealand (ASB, TPB), a Rutherford Discovery Fellowship administered by the Royal Society of New Zealand (ASB),  the University of Otago (MTR), and the US National Science Foundation [BPA, (PHY-1205713)].

\end{document}